\documentclass[a4paper,onecolumn,11pt]{article}

\usepackage{amsmath,amssymb}

\newcommand{\mcl}{\mathcal{L}}
\newcommand{\mcs}{\mathcal{S}}
\newcommand{\mbr}{\mathbb{R}}
\DeclareMathOperator{\diag}{diag}

\newcommand{\half}{\frac{1}{2}}
\newcommand{\quart}{\frac{1}{4}}

\begin{document}


\rightline{NIKHEF/2009-028}

\vspace{0.8cm}

\centerline{\LARGE Dynamics of the infinitely-thin kink}

\vspace{0.8cm}

\centerline{\large Damien P. George\footnote{\tt dpgeorge@nikhef.nl}}
\centerline{\large Nikhef, Science Park 105, 1098 XG Amsterdam, The Netherlands}
\centerline{\large School of Physics, The University of Melbourne, Victoria 3010, Australia}

\vspace{0.5cm}

\centerline{\large Raymond R. Volkas\footnote{\tt raymondv@unimelb.edu.au}}
\centerline{\large School of Physics, The University of Melbourne, Victoria 3010, Australia}

\vspace{0.5cm}

\centerline{\today}

\vspace{0.6cm}

\begin{abstract}

We consider the dynamics of the domain-wall kink soliton, in
particular we study the zero mode of translation.  In the infinitely-%
thin kink limit, we show that the zero mode is almost completely
frozen out, the only remnant being a dynamically constrained
four-dimensional mode of a single but arbitrary frequency.
In relation to this result, we show that the usual mode expansion for
dealing with zero modes --- implicit collective coordinates --- is
\emph{not} in fact a completely general expansion, and that one must
use instead a traditional generalised Fourier analysis.

\end{abstract}



\section{Introduction}

The classical kink soliton solution of the $\lambda\phi^4$ theory
has found many applications.  One such use has been in models of
extra-dimensions, where a background scalar field assumes the kink
solution and becomes a domain-wall brane, a specific realisation
of the generic idea of a brane world.  From the point of view of a
model builder, the kink can be used to localise fermions~%
\cite{Rubakov:1983bb, ArkaniHamed:1999dc}, gauge fields~%
\cite{Dvali:1996xe}, Higgs fields~\cite{George:2006gk} and gravity
\cite{Gremm:1999pj, Csaki:2000fc} (building on~%
\cite{Randall:1999vf}).  Giving the kink a non-trivial
representation under some internal symmetry allows for exciting
symmetry breaking opportunities, such as GUT breaking~%
\cite{Pogosian:2000xv, Vachaspati:2001pw, Pogosian:2001fm, Davidson:2002eu}
and supersymmetry breaking~\cite{Dvali:1996bg}.  All these
ingredients are able to play together in a comprehensive model of
extra-dimensions, and a domain-wall-localised standard model can
be implemented~\cite{Davies:2007xr}.

Even though the kink has played a central role in domain-wall
models for many decades now, there are some interesting and
important technical properties of the kink that have been
overlooked.  These loose ends were alluded to in a previous work
by the authors~\cite{George:2006gk}, and relate to the precise
nature of the zero mode of translation of the kink, the thin-kink
limit, and the implicit collective coordinate treatment.%
\footnote{For earlier analyses of the thin kink limit,
see~\cite{Gregory:1990pm, Carter:1994ag}.}
In this
paper we shall resolve these issues and make clear the following
two facts: first, that the kink zero mode corresponding to
translations is almost completely frozen out in the thin-kink
limit, and, second, that the implicit collective coordinate
expansion (ICCE) does not capture all physically-acceptable 
classical field configurations.  Both of these results appear to
be contrary to common understanding, they impact the conclusions
of previous work, and they must be taken into account in future
studies of kinks.

Our analysis is chiefly mathematical and the results are valid for
any application of the kink solution, not just domain-wall brane
theories.  But to aid in physical understanding and help the flow
of our argument, we have in mind the specific scenario of a five-%
dimensional theory with a bulk scalar field forming a kink.  We
are interested in integrating out the extra-dimension to determine
the equivalent four-dimensional theory, and we shall elucidate the
scalar degrees of freedom present in this reduced spacetime.  The
thin-kink limit is an important phenomenological limit for such a
model, as the masses of the Kaluza-Klein (KK) modes are pushed to
infinity.  In addition, the action for a thin kink can be compared
with the Nambu-Goto action for a fundamental brane.  For the case
of the infinitely-thin kink, we show that the Nambu-Goldstone
boson, related to the spontaneous breaking of the translation
symmetry, is not fully dynamical: the only remnant of translation
invariance in the four-dimensional theory is the allowance of a
single frequency massless mode.  When dealing with translation
invariance, one usually employs the ICCE; 
see Rajaraman~\cite{Rajaraman:1982is} and references
therein.  We shall demonstrate that such an expansion must be used
with caution, as it is not able to adequately encode all field
configurations of the original five-dimensional field and does not
properly handle the non-linear interactions of the zero-mode at
high order.

The paper is organised as follows.  In Section~\ref{sec:frozen} we
review the kink solution, its energy density and its behaviour in
the thin-kink limit.  We demonstrate the existence of a `wavy
kink' solution of a fixed frequency, which persists in the thin-%
kink limit.  We then argue that, in such a limit, this fixed
frequency wave is the only remaining dynamical behaviour and hence
the four-dimensional zero mode~--- the Nambu-Goldstone boson
corresponding to translations of the kink~--- is almost completely
frozen out.  In Section~\ref{sec:modes} we analyse the modes of
the kink, showing that the ICCE
is not completely general, and we use the fully-general
Fourier expansion to show that the zero mode is truly frozen out.
We make some further remarks regarding dimensional reduction and
then conclude in Section~\ref{sec:concl}.


\section{The `wavy kink' and the frozen zero mode}
\label{sec:frozen}

The set-up of the problem is quite simple; we consider five-%
dimensional Minkowski spacetime, and a single scalar field with a
quartic potential.  The action is
\begin{equation}
\label{eq:phi-act-5d}
\mcs = \int d^5x \left[ \half \partial^M \Phi \partial_M \Phi - V(\Phi) \right] \:,
\end{equation}
where $\Phi$ is the scalar field, and the potential is
\begin{equation}
V(\Phi) = \frac{\lambda}{4} \left( \Phi^2 - v^2 \right)^2 \:.
\end{equation}
Indices $M, N$ run over the spacetime coordinates $(t,x,y,z,w)$,
the Minkowski metric is $\eta_{MN}=\diag(+1,-1,-1,-1,-1)$, and $\lambda$
and $v$ are real parameters.  The equation of motion for $\Phi$ is
\begin{equation}
\label{eq:phi-el}
\partial^M \partial_M \Phi - v^2 \lambda \Phi + \lambda \Phi^3 = 0 \:.
\end{equation}

The well-known classical kink solution to Eq.~%
\eqref{eq:phi-el} is
\begin{equation}
\label{eq:phi-clas}
\phi_c(w) = v \tanh \left( k w \right) \:,
\end{equation}
where $k=v\sqrt{\lambda/2}$ is the inverse width of the kink.
Here, we have chosen the kink profile to depend on the extra-%
dimensional coordinate $w$, as this is the dimension we want to
eliminate when constructing the equivalent four-dimensional theory.
Integrating over $w$, one obtains the energy density per unit
four-volume of the kink:
\begin{equation}
\varepsilon = \int dw \left[ \half \phi_c'^2 + V(\phi_c) \right]
    = \frac{2}{3} v^3 \sqrt{2\lambda} \:.
\end{equation}

The thin-kink limit has the width of the hyperbolic tangent
profile tending to zero, and is defined by $k\to\infty$ while
$\varepsilon$ is kept finite.  For the two parameters of the
model, this limit translates to $\lambda\to\infty$ and $v\to0$,
with $v^6 \lambda$ finite.

We would now like to make a less restrictive ansatz for the
solution to the five-dimensional Euler-Lagrange equation, an
ansatz which can describe degrees of freedom on top of the static
kink profile.
Due to the Poincar\'e invariance of the action, any $w$-translated
form of Eq.~\eqref{eq:phi-clas} is also a valid solution
for $\Phi$.  Using this fact as a hint, we try the more general
translated ansatz
\begin{align}
\Phi(x^M) &= \phi_c\left(w-Z(x^\mu)\right) \\
\label{eq:phi-wavy}
          &= v \tanh \left[ k \left(w-Z(x^\mu)\right) \right] \:.
\end{align}
Here, the index $\mu$ runs over the four-dimensional subspace
$(t,x,y,z)$.  $Z(x^\mu)$ is a real scalar field which acts to
translate the kink by an $x^\mu$-dependent amount, and includes as
a particular case any constant shift of the kink.
This ansatz is in fact of the same form as the first
term in the ICCE approach to redescribing the five-dimensional
scalar field as an infinite tower of four-dimensional KK scalar
fields.  In terms of the ICCE, we have here taken the solutions
for all massive KK four-dimensional fields to be zero.  We shall
examine the general expansion, which retains all KK modes, in the
next section.

The fundamental theory is that of a five-dimensional scalar field.
To ensure that all of the physics is retained, the correct approach to
finding solutions for $Z(x^\mu)$ is therefore to substitute the ansatz into
the five-dimensional Euler-Lagrange equation~\eqref{eq:phi-el}.
Doing this gives 
\begin{equation}
\label{eq:wavy-el}
- \phi'_c(w-Z) \partial^\mu\partial_\mu Z + \phi''_c(w-Z) \partial^\mu Z \partial_\mu Z = 0 \:,
\end{equation}
where prime denotes derivative with respect to $w$.
Since $\phi_c''$ is an odd function, integrating this equation
over $w$ eliminates the second term,%
\footnote{Note that in doing the integration over $w$ we have
terms such as $\int \phi'_c(w-Z) dw$ which look like functions
of $x^\mu$.  These terms actually yield the same value for each
point in the four-space (which can be seen by changing the
integration variable independently at each $x^\mu$), and so the
integral results in an $x^\mu$-independent answer.}
and so the most general solution obeys
\begin{equation}
\label{eq:z-constrain}
\partial^\mu\partial_\mu Z = \partial^\mu Z \partial_\mu Z = 0 \:.
\end{equation}
Solutions for $Z(x^\mu)$ are massless plane waves \emph{of a
single frequency only}.  The usual equation of motion for a zero
mode, $\partial^\mu\partial_\mu Z=0$, now has an auxiliary
constraint, $\partial^\mu Z \partial_\mu Z = 0$, and one can no
longer take a Fourier sum of all frequencies.  The most general
solution to both of these equations is
\begin{equation}
\label{eq:z-soln}
Z(x^\mu) = A\cos(p_\mu x^\mu) + B\sin(p_\mu x^\mu) + C \:,
\end{equation}
with $A$, $B$ and $C$ arbitrary real numbers and $p_\mu p^\mu=0$.
Notice that this solution solves the five-dimensional equation of
motion~\eqref{eq:wavy-el} irrespective of the values of the
parameters $\lambda$ and $v$; in particular, it remains valid in
the thin kink limit.
The auxiliary constraint means that, as an effective four-%
dimensional field, $Z$ does not
manifest as a standard dynamical scalar field in the four-%
dimensional theory.  

Let us now compute the energy density per unit four-volume for the 
more general kink solution given by Eqs.~%
\eqref{eq:phi-wavy} and~\eqref{eq:z-constrain}. It is%
\footnote{As before, one will encounter terms such as
$\int \phi'^2_c(w-Z) dw$ which are actually $x^\mu$-independent.}
\begin{align}
E &= \int dw \left[ \half \dot{\Phi}^2
    + \half \left( \nabla\Phi\cdot\nabla\Phi \right)
    + \half \Phi'^2 + V(\Phi) \right] \\
    &= \varepsilon
    + \half \varepsilon \dot{Z}^2
    + \half \varepsilon \left( \nabla Z \cdot \nabla Z \right) \:.
\end{align}
Here, an over-dot denotes derivative with respect to $t$.  $E$ is
the energy density of the original kink background, $\varepsilon$,
plus kinetic and gradient energy of $Z$, with larger energy for
higher frequency $Z$ solutions.  The energy density is not
sensitive to the individual parameters $v$ and $\lambda$, only
their combination $\varepsilon$.  Importantly, in the infinitely-%
thin kink limit, we are allowed a non-zero form for $Z$, as its
contribution to the total energy density remains finite (assuming
the spacetime derivatives of $Z$ are finite).

Summarising, we have found a slightly more general kink solution,
given by Eq.~\eqref{eq:phi-wavy}, which is an \emph{exact}
solution of the five-dimensional Euler-Lagrange equation so long
as $Z$ takes the form of Eq.~\eqref{eq:z-soln}.  Due to the
fact that this solution for $Z$ must be of a fixed frequency (with
arbitrary phase and amplitude), we shall call the resulting
solution the `wavy kink' solution.  The `wave' appears along the
length of the kink such that the hyperbolic tangent profile is
shifted in the $w$-direction by an amount that varies sinusoidally
in the three-space $(x,y,z)$.  This wave oscillates in time at a
fixed frequency, and, from the point of view of a four-dimensional
observer, is the only dynamical behaviour that can be observed
given the ansatz~\eqref{eq:phi-wavy}.  Consequently, $Z$ cannot be
called a proper four-dimensional mode, as, from a momentum-space
perspective, its degrees of freedom consist of a set of measure
zero: a single frequency.

Now, it may seem that we have been too restrictive in our ansatz
for the scalar field.  After all, conventional wisdom has it that
the kink spontaneously breaks translation invariance, and so there
should be a massless Nambu-Goldstone boson at the effective four-%
dimensional level.  This boson would correspond to translations of
the kink.  In fact, at first order, this Nambu-Goldstone boson is
exactly $Z$: for small $Z$, where we ignore $Z^2$ and higher
terms, the auxiliary constraint in Eq.~\eqref{eq:z-constrain}
is eliminated.  In this approximation, we are left only with the
usual massless wave equation describing the behaviour of $Z$, and
so the system admits a fully-dynamical scalar mode at the four-%
dimensional level.

One may think that this Nambu-Goldstone zero mode should persist,
even if we move outside the regime of the approximation and keep
higher order terms for $Z$.  We shall show that this is not
actually the case, and that the necessary coupling of $Z$ to
higher-mass modes constrains its dynamics.  What follows is a
brief intuitive argument for such behaviour.  In the next section
we provide a more rigorous mathematical analysis.

Consider what happens if one excites the scalar field $\Phi$ to a
configuration $\Phi(w-Z)$ where the form of $Z$ consists of
multiple frequencies.  Obviously, such an excitation does not
satisfy the five-dimensional Euler-Lagrange equation~%
\eqref{eq:phi-el} with our restricted ansatz~\eqref{eq:phi-wavy}.
What will happen is that, as the system evolves in time, the
KK fields set to zero in our ansatz will be excited.
It is important to realise that if we prohibit the excitation of
such modes, then we can only have solutions of
the form $\Phi(w-Z)$ with $Z$ a single frequency massless plane
wave.  Now we are in a position to state one of the main results
of this paper: in the infinitely-thin kink limit, such extra modes
are infinitely heavy (they are frozen out), and so the zero mode
$Z$ is dynamically constrained to a single frequency.  As a
consequence, from the effective four-dimensional point of view,
$Z$ does not have enough degrees of freedom to look anything like
a traditional scalar field, and so this Nambu-Goldstone boson is
not present in the four-dimensional spectrum.  In fact, in the
infinitely-thin kink limit, the four-dimensional spectrum contains
no propagating degrees of freedom at all.


\section{Collective coordinates and the general expansion}
\label{sec:modes}

In this section we proceed to analyse the full spectrum of modes
of the kink, and demonstrate that all the propagating degrees of
freedom are frozen out in the thin-kink limit.  To do this, we
shall expand the five-dimensional field $\Phi$ in a set of
complete four-dimensional modes~--- a generalised Fourier
transformation, or Kaluza-Klein decomposition.  The extra
dimension can then be integrated out to obtain a four-%
dimensional action, giving an equivalent, but alternative,
description of the original theory.  The appropriate expansion is
written as
\begin{equation}
\label{eq:mode-expansion}
\Phi(x^\mu,w) = \phi_c(w) + \sum_i \phi_i(x^\mu) \eta_i(w) \:,
\end{equation}
where the sum over $i$ includes two discrete modes ($i=0,1$) and
an integral over a continuum ($i=q$, where $q\in\mbr$).  The
profiles $\eta_i(w)$ form a complete basis (in the sense that
\emph{any} physically-acceptable five-dimensional field 
configuration $\Phi(x^\mu,w)$ can be
represented by suitable choice of $\phi_i(x^\mu)$) and are
determined by linearising the five-dimensional Euler-Lagrange
equation about the kink background; see reference~%
\cite{George:2006gk} for explicit forms of the basis functions
$\eta_i$.  Note that even though the $\eta_i$ were determined after
linearising, they still form a complete basis in the exact regime,
and can be used for a general expansion with no loss of
information.  The fields $\phi_i$ are a tower of scalar fields,
and serve to faithfully represent, at the four-dimensional level,
all degrees of freedom inherent in $\Phi$.  The tower consists of
a zero-mass mode, followed by a discrete massive mode, followed by
a massive continuum.

Before using this expansion, we shall discuss a slightly different
version of Eq.~\eqref{eq:mode-expansion}, the aforementioned ICCE~%
\cite{Rajaraman:1982is}.  Since any translated version of the
background kink $\phi_c$ is just as good as any other, there
exists an entire class of basis functions $\eta_i$ which are also
translated by an equivalent amount.  The first basis function
$\eta_0$ is proportional to the first derivative of $\phi_c$ and
corresponds to infinitesimal (first order) translations of the
static kink profile.  The mode $\eta_0$ therefore plays a unique
role, and it should perhaps be treated differently from the other
$\eta_i$ modes.  The ICCE is motivated by this observation, and
removes the zero mode from the tower of modes, placing it in a
more `obvious' spot:
\begin{equation}
\label{eq:cc-expansion}
\Phi(x^\mu,w) = \phi_c\left(w-Z(x^\mu)\right)
    + \sum_{i\ne0} \tilde{\phi}_i(x^\mu) \eta_i\left(w-Z(x^\mu)\right) \:.
\end{equation}
The idea now is that the four-dimensional scalar fields
$Z(x^\mu)$ and $\tilde{\phi}_{1,q}(x^\mu)$ can faithfully encode
all degrees of freedom of $\Phi$.  Note that, in this expansion,
$\phi_c$ and $\eta_{1,q}$ have the same form as they
do in Eq.~\eqref{eq:mode-expansion}, but now the sum excludes
$i=0$.  The ICCE has seen numerous applications to problems
where continuous symmetries and zero modes are present.  For
example, in a perturbative quantum field theory analysis, the zero
mode can potentially lead to divergent energy contributions in
higher-order terms~\cite{Rajaraman:1982is}.  The ICCE allows one
to treat the zero mode separately and avoid such difficulties.

Although Eq.~\eqref{eq:cc-expansion} looks quite reasonable,
it is actually not general enough to expand an arbitrary field
$\Phi(x^\mu,w)$.  For example, there are no (finite) choices of
$Z(x^\mu)$ and $\tilde{\phi}_{1,q}(x^\mu)$ which yield
$\Phi(x^\mu,w)=\omega(x^\mu)\eta_0(w)$ for \emph{any} non-zero
choice for $\omega(x^\mu)$.%
\footnote{Note that it is not necessary for the configuration
$\omega(x^\mu)\eta_0(w)$ to be a classical solution.  It is enough that it exists in the
space of all possible configurations.  At the level of the action, the field $\Phi$
is, of course, taken to be a variable and this configuration is one possible `value'
this variable can take.  At the quantum level, the path integral must include this
configuration in the domain of functional integration.}  
If there were, then we could write
\begin{equation}
\label{eq:cc-proof-1}
\omega(x^\mu)\eta_0(w) = \phi_c(w-Z) + \sum_{i\ne0} \tilde{\phi}_i(x^\mu) \eta_i(w-Z) \:.
\end{equation}
Keep in mind that $Z$ may depend on $x^\mu$, we have just
neglected to write this explicitly to keep the equation clear.
Now, multiply through by $\eta_0(w-Z)$ and integrate over $w$:
\begin{equation}
\begin{aligned}
\omega(x^\mu) \int \eta_0(w)\eta_0(w-Z) \; dw
    &= \int \phi_c(w-Z)\eta_0(w-Z) \; dw \\
    &\quad+ \sum_{i\ne0} \tilde{\phi}_i(x^\mu) \int \eta_i(w-Z)\eta_0(w-Z) \; dw \:.
\end{aligned}
\end{equation}
There is the freedom to shift the integrals on the righthand-side
by $Z$, and then, because $\eta_0$ is orthogonal to $\phi_c$ and
$\eta_{1,q}$, we have
\begin{equation}
\label{eq:cc-proof-2}
\omega(x^\mu) \int \eta_0(w)\eta_0\left(w-Z(x^\mu)\right) \; dw = 0 \:,
\end{equation}
Since $\eta_0$ is strictly positive (or strictly negative,
depending on the normalisation convention) the integral in this
equation will always be positive, regardless of the form of
$Z(x^\mu)$, and so it must be that $\omega(x^\mu)=0$.
(We shall discuss shortly the possibility that $Z$ is infinite).
Hence we have shown that the implicit
collective coordinate expansion~\eqref{eq:cc-expansion} cannot
faithfully represent all possible configurations of $\Phi$, and so is less
general than the mode expansion~\eqref{eq:mode-expansion}.

We should make clear what we mean by an expansion being general
enough to represent any (physically-acceptable) classical field 
configuration.  In one-dimensional, non-relativistic quantum mechanics, one looks for
the eigenfunctions of a time-independent Schr\"odinger equation,
and builds a set out of those eigenfunctions which are
\emph{bounded at infinity}.  Relying on Sturm-Liouville theory, one
can make the statement that this set forms a complete set of modes,
and any function that is also bounded at infinity can be expanded as
a linear combination of the eigenfunctions.  It is this idea of
completeness that we have in mind throughout the current paper.
Our argument above demonstrates that the ICCE is not general enough
to represent an arbitrary configuration which is bounded at
infinity.  In contrast, the mode expansion given by Eq.%
~\eqref{eq:mode-expansion} is determined from a Schr\"odinger-like
equation as the set of eigenfunctions which are bounded at infinity,
and so is able to represent a more general, and in fact adequate,
class of configurations than the ICCE.%
\footnote{We have checked explicitly that the expansion~\eqref{eq:mode-expansion}
can represent the configuration $\Phi(x^\mu,w)=\omega(x^\mu)\eta_0(w)$
for any choice of $\omega(x^\mu)$.  Essentially, there exists a linear
combination of the massive modes $\eta_{1,q}(w)$ which cancel the
kink configuration $\phi_c(w)$, leaving the zero mode $\eta_0(w)$.}

In an attempt to satisfy Eq.~\eqref{eq:cc-proof-2}, one
may try and take $Z(x^\mu)\to\infty$, in which case the overlap of
the two $\eta_0$ profiles becomes infinitesimally small and the
integral vanishes~\cite{Burnier:2009}.
If we allow $Z$ to be infinite, our argument
above (that the ICCE is not general) breaks down because
multiplying Eq.~\eqref{eq:cc-proof-1} through by
$\eta_0(w-Z)$ is essentially multiplying through by zero.
To understand what is happening, must look back at the actual
definition of the ICCE, Eq.~\eqref{eq:cc-expansion}, and
consider the effect of taking $Z\to\infty$.  Mathematically,
the $\phi_c(w-Z)$ term becomes a constant ($+v$ or $-v$, depending
on whether $Z\to-\infty$ or $Z\to\infty$, respectively), the
discrete mode $\eta_1(w-Z)$ vanishes, and the continuum modes
$\eta_q(w-Z)$ become plane waves with frequency beginning at zero.
In such a limit, the ICCE thus reduces to the standard Fourier
transform of sines and cosines.  Physically, one has translated
the kink away, off to infinity, leaving a homogeneous vacuum
(which is of course an allowed solution of the theory).
Using the ICCE
with $Z$ of infinite magnitude is therefore equivalent to doing a
mode decomposition around a homogeneous vacuum background,
rather than around the kink background.  Although the underlying
five-dimensional theory can be equally-well recast into
equivalent four-dimensional forms using either decomposition (around
the kink background, or around a homogeneous vacuum), for a 
given application one re-description will be more convenient than the
other.  For the application of concern to us here, the kink-background
approach is clearly the more convenient.  If one were to try to do
the analysis using the homogeneous-vacuum mode (standard Fourier)
basis, one would first have to understand how to choose
the Fourier coefficients to produce the kink background plus
excitations.  This can be done, of course, but it is an awkward 
way to proceed.  So, with a finite $Z$ the ICCE is not fully general, while
for infinite $Z$ one recovers a mode basis that is not convenient
for studying kink-related physics.

There are certain regimes of analysis where the ICCE is adequate.
This includes the case where one restricts oneself to look only at
small perturbations of the kink background, as there are no troubles
expanding a perturbed kink using the ICCE.  The configuration used
in the above argument~--- the one which cannot be represented by the
ICCE, $\Phi(x^\mu,w)=\omega(x^\mu)\eta_0(w)$~--- is not a small
perturbation of the kink since its asymptotic behaviour differs from
that of $\phi_c(w)$.  In this paper we are interested in determining
the full, non-linear behaviour of the kink \emph{exactly}, and must
allow for the possibility that the kink background configuration is
significantly modified.  The ICCE approach is therefore unsuitable.
Instead, using the more general expansion~\eqref{eq:mode-expansion}
allows us to transform the five-dimensional kink into an equivalent
description in terms of four-dimensional scalar modes.  Analysing
these modes is then straightforward because the resulting Lagrangian
contains just massive interacting fields, with usual Klein-Gordon
equations of motion (as opposed to the ICCE which yields
difficult-to-interpret derivative couplings).  This is due to the
proper choice of basis functions $\eta_i$.%
\footnote{At the very least, regardless of the regime of validity
of the ICCE, the mode expansion given by Eq.~%
\eqref{eq:mode-expansion} is general enough to describe all
configurations which remain bounded at infinity, and we can be
confident in using it to obtain an equivalent four-dimensional
action.  If the ICCE is equally valid, it should produce the
same results.}

Substituting Eq.~\eqref{eq:mode-expansion} in the five-%
dimensional action~\eqref{eq:phi-act-5d} and integrating out the
extra dimension yields the equivalent four-dimensional action:
\begin{equation}
\mcs_\Phi = \int d^4x \left[ -\varepsilon_{\phi_c} + \mcl_\phi \right],
\end{equation}
where the kinetic, mass and self-coupling terms for the scalar
modes are (see~\cite{George:2006gk})
\begin{equation}
\begin{aligned}
\label{eq:phi-lag-4d}
\mcl_\phi &=
    \half \partial^\mu \phi_0 \partial_\mu \phi_0
    + \half \partial^\mu \phi_1 \partial_\mu \phi_1
    - \frac{3}{4} v^2 \lambda \phi_1^2 \\
& \quad
    + \int_{-\infty}^{\infty} dq \left[
        \half \partial^\mu \phi_q^* \partial_\mu \phi_q
        - \quart (q^2 + 4) v^2 \lambda \phi_q^* \phi_q \right] \\
& \quad
    - \kappa^{(3)}_{i j k} \phi_i \phi_j \phi_k
    - \kappa^{(4)}_{i j k l} \phi_i \phi_j \phi_k \phi_l \:.
\end{aligned}
\end{equation}
In the last line here there are implicit sums over discrete modes,
and integrals over continuum modes, for each of the indices $i$,
$j$, $k$ and $l$.  For these terms, the cubic and quartic coupling
coefficients are, respectively,
\begin{align}
\kappa^{(3)}_{i j k} &= \lambda \int_{-\infty}^\infty \phi_c \eta_i \eta_j \eta_k \; dw \:, \\
\kappa^{(4)}_{i j k l} &= \frac{\lambda}{4} \int_{-\infty}^\infty \eta_i \eta_j \eta_k \eta_l \; dw \:.
\end{align}

The four-dimensional equivalent theory described by Eq.~%
\eqref{eq:phi-lag-4d} contains a massless scalar field $\phi_0$,
a massive scalar $\phi_1$, and a continuum of massive fields
$\phi_q$.  There exist cubic and quartic couplings among these
fields, and, importantly, a quartic \emph{self-coupling} term for
$\phi_0$; this is due to the non-zero value of
$\kappa^{(4)}_{0000}$:
\begin{equation}
\kappa^{(4)}_{0000} = \frac{9}{70} \left(\frac{3\varepsilon}{8}\right)^{1/3} \lambda^{4/3} \:.
\end{equation}
Determining the Euler-Lagrange equations for each of the fields is
a straightforward task.  For our purposes, it suffices to examine
the two discrete modes:
\begin{align}
\label{eq:phi0-el}
\partial^\mu\partial_\mu \phi_0
    + 6 \kappa^{(3)}_{001} \phi_0 \phi_1 
    + 4 \kappa^{(4)}_{0000} \phi_0^3
    + 12 \kappa^{(4)}_{0011} \phi_0 \phi_1^2 \nonumber\\
    + \; \text{(terms involving continuum modes)} &= 0 \:,\\
\label{eq:phi1-el}
\partial^\mu\partial_\mu \phi_1
    + \frac{3}{2} v^2 \lambda \phi_1
    + 3 \kappa^{(3)}_{001} \phi_0^2
    + 3 \kappa^{(3)}_{111} \phi_1^2
    + 12 \kappa^{(4)}_{0011} \phi_0^2 \phi_1
    + 4 \kappa^{(4)}_{1111} \phi_1^3 \nonumber\\
    + \; \text{(terms involving continuum modes)} &= 0 \:.
\end{align}

Given this rather neat, and exact, dimensional reduction of the
original $\Phi$ model, we can now make a rigorous conclusion
regarding the thin-kink limit.  In this limit,
$v^2 \lambda \to \infty$ and so the mass terms in the Lagrangian,
Eq.~\eqref{eq:phi-lag-4d}, become infinitely large.  From the
point of view of the Euler-Lagrange equations for $\phi_1$ and
$\phi_q$, the mass terms for these fields have an infinite
coefficient, and these equations of motion can only be generally
satisfied if the associated fields are identically zero.  We
therefore conclude that, in the infinitely-thin kink limit, the
massive modes $\phi_1$ and $\phi_q$ are frozen out.

Since $\phi_1$ and the continuum modes must be zero, Eq.~%
\eqref{eq:phi1-el} reduces to $3\kappa^{(3)}_{001}\phi_0^2=0$,
implying that $\phi_0$ must also be zero.%
\footnote{Analysis of the Euler-Lagrange equations for the
continuum modes reveals similar constraints, such as
$\kappa^{(3)}_{00q}\phi_0^2=0$ ($q$ corresponding to an odd mode)
and $\kappa^{(4)}_{000p}\phi_0^3=0$ ($p$ corresponding to an even
mode).}
This is the central part of the argument, and supports our earlier
claim that $Z$ is constrained due to its coupling to massive,
frozen modes.  Here, the dynamics dictate that $\phi_0$ must
excite $\phi_1$ (if $\phi_1$ begins as zero) and so if $\phi_1$ is
forbidden (for example, if it is infinitely heavy), then $\phi_0$
cannot be excited at all.  Similar statements can be made
regarding the coupling of $\phi_0$ to the massive continuum modes.
Furthermore, the quartic coupling of $\phi_0$ to itself also
prevents it from being excited: in the thin kink limit,
$\kappa^{(4)}_{0000}\to\infty$, and, in order to satisfy
Eq.~\eqref{eq:phi0-el}, $\phi_0$ is driven to zero.

From a slightly different point of view, consider all fields
$\phi_i$ to be identically zero to begin with, and attempt to
excite them individually.  In the thin kink limit, all of the
Euler-Lagrange equations contain potential terms that are infinite
if any one of the fields are independently excited.  In the
equation for $\phi_0$, this term has coefficient
$4\kappa^{(4)}_{0000}$, for $\phi_1$ it has
$\tfrac{3}{2}v^2\lambda$, and for $\phi_q$ it has
$\tfrac{1}{2}(q^2+4)v^2\lambda$.  Thus, each field is individually
frozen.

We are essentially arguing that, in the Euler-Lagrange equations
for the four-dimensional fields (and also in the four-dimensional
action), there are coefficients which become infinite in the thin
kink limit, and so the fields that make up such terms must be
zero at the solution level.  The reader may wonder if there exists
some special combination of these fields which conspire to cancel
the infinities.  This is actually true.  The special combination
of $\phi_0$ and the massive modes that persists in the thin-kink
limit is the fixed frequency, wavy kink solution that we found in
Section~\ref{sec:frozen}.  But this is not a true four-dimensional
dynamical field.  In reference~\cite{George:2006gk} it is shown
that there is no other special combination that manifests as a
proper four-dimensional scalar field with a canonical kinetic
term.

The mode expansion given by Eq.~\eqref{eq:mode-expansion}
retains all degrees of freedom of $\Phi$, and our analysis shows
that these degrees of freedom are all driven to zero in the thin
kink limit.  Furthermore, there is no special combination of the
modes which yields a field with a proper kinetic term.  We have
therefore shown that there are no observable dynamics, at the
four-dimensional level, of the infinitely-thin kink.  It is
perhaps best, then, to consider a thin kink as also being a rigid
kink; that is, it cannot be perturbed.  For a thin kink (but not
infinitely thin), one can set up a finite-energy-density
configuration $\phi_c(w-Z(x^\mu))$ with arbitrary form for
$Z(x^\mu)$.  But, if the kink is made thinner, and hence more
rigid, the same configuration will have a greater energy cost and
will dissipate more rapidly to a wavy kink of fixed frequency
(possibly zero frequency: the usual, static kink).  For the case
of the infinitely-thin kink limit, the initial configuration must
\emph{begin} as a fixed frequency wavy kink.


\section{Discussion and conclusion}
\label{sec:concl}

It must be stressed again that a five-dimensional theory is ruled by the
five-dimensional Euler-Lagrange equations.  Four-dimensional
Euler-Lagrange equations provide an equivalent description only
when all five-dimensional fields have been expanded in a full, or
general, set of modes.  To go to an equivalent four-dimensional
theory, one should \emph{not} make a non-general ansatz for the five-%
dimensional fields, then integrate out the extra dimension.
To find a four-dimensional theory which is equivalent to the
original five-dimensional one, all degrees of freedom must be kept
to begin with, and then the irrelevant ones eliminated at the
four-dimensional level.

If one does not begin with a general expansion of the five-%
dimensional fields, then one may miss some important low-energy
dynamics, dynamics which influence the behaviour of other low-%
energy degrees of freedom that have been included.  For a concrete
example of this statement, consider the non-general expansion
$\Phi=\phi_c(w-Z(x^\mu))$.  There is nothing wrong with employing
such an expression as a solution ansatz, but, since it ignores a
great number of degrees of freedom, one must use the five-%
dimensional Euler-Lagrange equation to determine the behaviour of
$Z$.  This is what we did in Section~\ref{sec:frozen}, where we
found that $Z$ must be a massless plane wave of a fixed frequency.
Now, to contrast this method, we try and substitute the non-%
general expansion into the original five-dimensional action,
integrate out the extra dimension, and obtain the effective four-%
dimensional action:
\begin{align}
\mcs &= \int d^5x \left[
        \half \phi_c'^2(w-Z) \: \partial^\mu Z \partial_\mu Z
        - \half \phi_c'^2(w-Z)
        - V\left(\phi_c(w-Z)\right) \right ] \\
    &= \int d^4x \left[
        \half \varepsilon \: \partial^\mu Z \partial_\mu Z
        - \varepsilon \right] \:.
\end{align}
This procedure gives the four-dimensional Euler-Lagrange equation
$\partial^\mu\partial_\mu Z=0$, which is not correct, as it is
missing the auxiliary constraint that fixes $Z$ to a single
frequency.  The first method we used is the correct method, as the
solution respects the full five-dimensional theory.  Reduction to
lower dimensions can only proceed if one uses a full, general mode
expansion.

In light of this argument, there is no sense in using the ICCE
to redescribe a five-dimensional theory as a completely equivalent four-%
dimensional theory.  As we have shown, the ICCE is not general,
and, in going to a four-dimensional description, one will
potentially miss out on degrees of freedom which are pertinent to
the low-energy dynamics.

Having said this, the ICCE is useful in certain contexts; for
example, where one is only interested in expanding a model up to a
given order in perturbation theory.  This is actually the case for
the discussions in Rajaraman~\cite{Rajaraman:1982is}, where modes
are quantised around a classical ground state (like the kink), and
perturbation theory is used to analyse the quantum excitations.
As pointed out in Section~\ref{sec:frozen}, if one works in the
regime where $Z$ is small and $Z^2\sim0$, the auxiliary constraint
$\partial^\mu Z \partial_\mu Z=0$ is automatically satisfied at
this order, and one is allowed a fully dynamical field $Z(x^\mu)$
at the four-dimensional level.  Physically, this means that $Z$ is
so small that it does not excite the higher-mass modes.

As a relevant aside, we shall make some brief comments regarding
fundamental branes.
Such branes may originate from string theory, and
are modelled by an effective action~--- the Nambu-Goto action~---
which treats them as infinitely thin, delta-distribution sources.
These branes are assumed to support a proper translation zero mode
which couples only through derivative terms to other fields (via
the metric).  We can accept this behaviour by understanding that
branes modelled by the Nambu-Goto action are flexible, even though
they are infinitely thin.  Their degree of flexibility is dictated
by their tension, which is equivalent to their energy density.  In
contrast, modelling a brane by a thin domain-wall kink solution
yields different effective four-dimensional dynamics; the domain
wall does not exhibit a dynamical zero mode, as it is extremely
rigid.

Our conclusion that the ICCE is not a general expansion may
have an impact on previous work that relied on this method.
For example, Burnier and Zuleta~\cite{Burnier:2008ke} compared
fundamental branes and kinks using the ICCE for two scalar fields
(the kink and an additional coupled scalar).  The low-energy
expansion of the domain-wall model was compared with the low-%
energy expansion of the Nambu-Goto action.  Their use of the ICCE
to describe \emph{perturbations} of the kink is well justified,
but it is not clear to us that their conclusions would remain
unchanged using the more general mode expansion given by
Eq.~\eqref{eq:mode-expansion}.  Another
interesting analysis to revisit is that where gravity is
included~\cite{Shaposhnikov:2005hc}.  Here, the zero mode of the
kink mixes with gravitational degrees of freedom.  It would be
important to understand which effects are more important: particle
physics modifications to the zero mode due to its quartic self-%
coupling, or the mixing with gravity.

In conclusion, we have established two facts that have been
overlooked during the study of domain walls and of kinks.  First,
that the zero mode of translation is almost completely frozen out
in the thin-kink limit.  The only remnant is a four-dimensional
entity which must assume a single frequency, yielding a wavy kink
solution.  This entity does not manifest as a proper mode in the
effective four-dimensional theory; almost all degrees of freedom
are frozen out.  Second, that the implicit collective coordinate
expansion is not completely general.  It should only be used with
caution and in certain approximations.


\section*{Acknowledgements}

We thank Y.~Burnier and K.~Zuleta for useful correspondence.
DPG was supported in part by the Puzey Bequest to the University
of Melbourne.  RRV was supported in part by the Australian
Research Council.



\end{document}